\documentclass[12pt,a4paper]{article}

\usepackage{a4}
\usepackage{amsfonts}
\usepackage{amssymb}
\usepackage{epsfig}
\usepackage{psfig}
\usepackage{psfrag}
\usepackage{dsfont}

\newcommand{\beqn}{\begin{eqnarray}}
\newcommand{\eeqn}{\end{eqnarray}}
\newcommand{\be}{\begin{equation}}
\newcommand{\ee}{\end{equation}}
\newcommand{\non}{\nonumber \\}

\begin{document}

\title{}
\begin{flushright}
\vspace{-3cm}
{\small MIT-CTP-3471 \\
        NUB-TH-3246 \\
        hep-ph/0402047
 }
\end{flushright}
\vspace{.5cm}

\begin{center}
{\Large\bf   A Stueckelberg Extension of the Standard Model}
\end{center}

\vspace{0.5cm}

\author{}
\date{}
%\maketitle
\thispagestyle{empty}

\begin{center}

{\bf Boris K\"ors}\footnote{e-mail: kors@lns.mit.edu}$^{,*}$
{\bf and Pran Nath}\footnote{e-mail: nath@neu.edu}$^{,\dag}$
\vspace{.5cm}

\hbox{
\parbox{8cm}{
\begin{center}
{\it
$^*$Center for Theoretical Physics \\
Laboratory for Nuclear Science \\
and Department of Physics \\
Massachusetts Institute of Technology \\
Cambridge, Massachusetts 02139, USA \\
}
\end{center}
}
\hspace{-.5cm}
\parbox{8cm}{\begin{center}
{\it
$^\dag$Department of Physics \\
Northeastern University \\
Boston, Massachusetts 02115, \\ USA \\
}
\end{center}
}
}

\vspace{.5cm}
\end{center}

\begin{center}
{\bf Abstract} \\
\end{center}
An extension of the standard model of electro-weak interactions by
an extra abelian gauge boson is given, in which the extra gauge
boson and the hypercharge gauge boson both couple to an axionic scalar in
a form that leads to a Stueckelberg mass term.  The theory leads
to a massive Z$'$ whose couplings to fermions are uniquely determined 
and suppressed by small mixing angles. Such a Z$'$ could have a low mass  
and appear in $e^+e^-$ collisions as a sharp resonance. 
The branching ratios into $f\bar f$ species, and the forward-backward 
asymmetry are found to have distinctive features.
The model also predicts a new unit of electric charge $e'=Q'e$, 
where $Q'$ is in general irrational, 
in the coupling of the photon with hidden matter that is neutral under 
$SU(2)_L\times U(1)_Y$.\\[1cm]

%\clearpage
\setcounter{footnote}{0}

The Stueckelberg mechanism \cite{stueck} gives mass to abelian
vector bosons without breaking gauge invariance on the Lagrangian,
and thus provides an alternative to the Higgs mechanism
\cite{higgs} to achieve gauge symmetry breaking without spoiling
renormalizability. The simplest case is that of one abelian vector
boson $A_\mu$  coupling to one axionic scalar field $\phi$. Here
the Lagrangian
\beqn
{\cal L} = -\frac14 {\cal F}_{\mu\nu}{\cal F}^{\mu\nu} - \frac{m^2}{2} A_\mu A^\mu
\eeqn
is made gauge invariant by splitting off the longitudinal degree of freedom
$\phi$ through $A_\mu \rightarrow A_\mu + \frac{1}{m}
\partial_\mu \phi$, and defining the gauge transformation $\delta
A_\mu = \partial_\mu \epsilon $, $\delta \phi = - m \epsilon$.
Thus, $\phi$ takes the role of the longitudinal component of the
massive vector, and is subject to a Peccei-Quinn type shift
symmetry, which arises for constant $\epsilon$. The interaction
with fermions is described by the standard (renormalizable)
coupling to a conserved current ${\cal L}_{\rm int} = g A_\mu
J^\mu$, $\partial_\mu J^\mu = 0$. In the quantum theory, a gauge
fixing term similar to the $R_\xi$ gauge is added to the
Lagrangian ${\cal L}_{\rm gf} = - (2\xi)^{-1}
\left( \partial_\mu A^\mu + \xi m \phi \right)^2$, such that the total Lagrangian reads
\beqn
\hspace{-.2cm}
{\cal L}+{\cal L}_{\rm int}+{\cal L}_{\rm gf} =
-\frac14 {\cal F}_{\mu\nu}{\cal F}^{\mu\nu} - \frac{m^2}{2} A_\mu A^\mu + 
g A_\mu J^\mu
- \frac{1}{2\xi} (\partial_\mu A^\mu)^2
%\non
%&&
- \frac12 \partial_\mu \phi \partial^\mu \phi - \xi \frac{m^2}{2}
\phi^2
\nonumber
\eeqn
where the two fields have been decoupled, and renormalizability
and unitarity are manifest. These ideas cannot be
easily extended to the non-abelian case \cite{nonabelian}, because
in the non-abelian extension of the Stueckelberg Lagrangian the
longitudinal components of the vector fields cannot be decoupled
from the physical Hilbert space, which spoils renormalizability
and unitarity. For the mass of the non-abelian gauge fields in the
standard model, a spontaneous symmetry breaking mechanism
involving the Higgs phenomenon \cite{higgs} is then
required.\footnote{The abelian Stueckelberg type of coupling among
gauge bosons and axionic scalars appear frequently in models of
(gauged) supergravity or string theory compactifications. The
``anomalous'' gauge transformation of the axions plays an
important role in the (generalized) Green-Schwarz anomaly
cancellation mechanism. In the framework of higher dimensional
Kaluza-Klein models, as arise from string theory, the
four-dimensional abelian vector field may be part of a larger
non-abelian, and possibly simple gauge group, that is broken in
the compactification.}

We now want to explore the perspectives of a model that extends
the standard model Lagrangian \cite{gws} by an abelian vector
boson and Stueckelberg type couplings. To start with, we look at
the relevant part of the standard model first. Let $A_{\mu}^a$ be
the gauge bosons in the adjoint of $SU(2)_L$ with field strengths
$F_{\mu\nu}^a$, $B_{\mu}$ the hypercharge $U(1)_Y$ vector with
field strength $B_{\mu\nu}$, and $\Phi$ be the $SU(2)_L$ Higgs
doublet. Then the relevant part of the standard model is given by
\beqn \label{sm}
{\cal L}_{\rm SM}= -\frac{1}{4} F_{\mu\nu}^aF^{a\mu\nu} -\frac{1}{4} B_{\mu\nu}B^{\mu\nu}
+ g_2 A^a_\mu J^{a\mu}_2 + g_Y B_\mu J^\mu_Y
- D_{\mu}\Phi^{\dagger} D^{\mu}\Phi - V(\Phi^{\dagger}\Phi)\ ,
\eeqn
where $D_{\mu}\Phi$ is the gauge covariant derivative and
$V(\Phi^{\dagger}\Phi)$ takes its minimum at $v^2/2$ as usual. For
the minimal extension of (\ref{sm}) by a Stueckelberg Lagrangian,
we add the degrees of freedom of one more abelian vector field
$C_\mu$ for a $U(1)_X$ with field strength $C_{\mu\nu}$, and one
axionic scalar $\sigma$.
 
We assume the scalar field $\sigma$ to have Stueckelberg couplings
to all the abelian gauge bosons, i.e. $B_\mu$ and $C_\mu$.
However, we leave untouched the charged vector bosons and the
charge assignment of the standard model fermions and assume they
are neutral under $U(1)_X$.\footnote{This important ingredient to
our model is naturally realized if $C_\mu$ belongs to a hidden
sector, and there are no fields except the axion $\sigma$ 
charged under both the hidden and the 
visible standard model gauge groups. Such an assumption would not
appear as natural, if we were generating masses only through
$SU(2)_L$ Higgs doublets (and maybe singlets), since then $C_\mu$
would be forced to be couple to $SU(2)_L\times U(1)_Y$ charges.}
Thus, the Lagrangian of Eq.(\ref{sm}) is extended by
\beqn
{\cal L}_{\rm St} = -\frac{1}{4} C_{\mu\nu}C^{\mu\nu}
+ g_X C_\mu J^\mu_X
- \frac{1}{2} (\partial_{\mu}\sigma  + M_1 C_{\mu} + M_2 B_{\mu})^2\ , 
\eeqn
where $J^{\mu}_X$ is the (conserved) hidden matter current, not
involving any standard model fields. The gauge invariance now
reads $\delta_Y B_{\mu} =
\partial_{\mu}\epsilon_Y$, $\delta_Y \sigma$ = $- M_2 \epsilon_Y$
for the hypercharge, and $\delta_X C_{\mu} =
\partial_{\mu}\epsilon_X$, $\delta_X \sigma = - M_1 \epsilon_X$
for $U(1)_X$. To decouple the various gauge bosons from the
scalars, one has to add similar gauge fixing terms as discussed
above (see e.g.\ \cite{Kuzmin:pg}). With the above extension, and
after the standard spontaneous electro-weak symmetry breaking the
mass terms in the vector boson sector take the form 
$-\frac12 V_{a\mu} M^2_{ab} V^\mu_b$, using $(V_\mu^{\rm
T})_a = ( C_{\mu}, B_{\mu}, A_{\mu}^3)_a$, with mass matrix
\beqn
M^2_{ab} =
\left(\matrix{ M_1^2  &  M_1M_2  &  0\cr
M_1M_2 & M_2^2 + \frac{1}{4} g_Y^2 v^2 & - \frac{1}{4}g_Yg_2 v^2
\cr 0 & -\frac{1}{4}g_Yg_2 v^2 & \frac{1}{4}g_2^2 v^2
}\right)_{ab}\ ,
\eeqn
where $v=2M_{\rm W}/g_2=(\sqrt 2 G_F)^{-\frac{1}{2}}$, $g_2$ and
$g_Y$ are the $SU(2)_L\times U(1)_Y$ gauge coupling constants,
$M_{\rm W}$ is the mass of the W boson, and $G_F$ is the Fermi
constant. From det$(M^2)=0$ it is easily seen that one eigenvalue
is zero, whose eigenvector we identify with the photon
$A_\mu^\gamma$. Among the remaining two eigenvalues $M_{\pm}^2$,
we identify the lighter mass eigenstate with mass $M_-$ with the
Z boson, and the heavy eigenstate with mass $M_+$ as the
Z$'$ boson. The limit $M_2\rightarrow 0$ takes one to the standard
model with $M_-^2=\frac14 v^2(g_2^2+g_Y^2)+{\cal O}(M_2^2)$ and
$M_+^2=M_1^2+{\cal O}(M_2^2)$. We pause here briefly to note that
in our analysis we succeeded in obtaining a massless photon,
whereas a previous attempt to obtain a Stueckelberg extension of
the standard model failed for that reason \cite{Kuzmin:pg}. It is
evident from counting degrees of freedom, that only two out of the
three gauge bosons $( C_{\mu}, B_{\mu}, A_{\mu}^3)$ can get
massive by absorbing the two scalars that are available.

Using an orthogonal transformation $O$ to diagonalize $M^2$,
 we go to the basis of eigenstates
$E^T=({\rm Z}_{\mu}',{\rm Z}_{\mu},A_{\mu}^\gamma)$, where
$V={O}\cdot E$, ${O}^T \cdot M^2 \cdot {O}=M^2_{\rm D}$, and
$M^2_{\rm D}={\rm diag}(M_{{\rm Z}'}^2, M_{\rm Z}^2,0)$. For ${O}$
we use the parametrization
\beqn %\label{A}
{O}=
\left[\matrix{ \cos\psi \cos\phi -\sin\theta\sin\phi\sin\psi &
-\sin\psi \cos\phi -\sin\theta\sin\phi\cos\psi & -\cos\theta \sin\phi\cr
\cos\psi \sin\phi +\sin\theta\cos\phi\sin\psi &
-\sin\psi \sin\phi +\sin\theta\cos\phi\cos\psi & \cos\theta \cos\phi\cr
-\cos\theta\sin\psi & -\cos\theta\cos\psi & \sin\theta }\right]
%\ .
\nonumber
\eeqn
The mixing angles $\theta$, $\phi$ and $\psi$ are given by\footnote{Note that $\theta$ and $\phi$ only  
involve the ratio $M_2/M_1$, such that the overall scale drops out. 
This is tempting in the framework of string or supergravity models, since 
it may lead to observable effects even for much higher mass scales than we consider in this paper.} 
\beqn
%\label{taninv}
\tan (\theta) = \frac{g_Y}{g_2}\cos (\phi)\ , ~
\tan (\phi) = \frac{M_2}{M_1} \ , ~
\tan (\psi) =
\frac{\tan(\theta)\tan(\phi)M_{\rm W}^2}
     {\cos(\theta)(M^2_{{\rm Z}'}-(1+\tan^2(\theta))M_{\rm W}^2)}\
     .
%\frac{g_1g_2v^2 \sin\phi}
%{(4M_{Z'}^2-g_2^2 v^2)\cos\theta -g_Yg_2\cos\phi \sin\theta}\ .
\nonumber
\eeqn
In the limit $M_2 \rightarrow 0$ one has $ \phi,\psi\rightarrow 0$
and $\theta\rightarrow\theta_W$, $\theta_W$ being the Weinberg
angle. 

The couplings to the physical vector fields in $E$ are
gotten by inserting the mass eigenstates into the interaction
Lagrangian
%
%\beqn
${\cal L}_{\rm int} = g_2 A_\mu^a J^{a\mu}_2 + g_Y B_\mu J^\mu_Y +
g_X C_\mu J^\mu_X.$
%\eeqn
%
We find the coupling of the photon given by
%
%\beqn
$A_\mu^\gamma ( eQJ^\mu_{\rm em} -e'J^{\mu}_X)$,
%\eeqn
where $e$ is the electric charge defined by
\beqn
e = \frac{g_2g_Y\cos(\phi)}{\sqrt{g_2^2+g_Y^2 \cos^2(\phi)}} \ .
\label{charge}
\eeqn
while $Q J^\mu_{\rm em} = \left( T_3 + \frac{Y}{2} \right)
J^\mu_{\rm em} = J_Y^\mu + J_2^{3\mu}$, and $e'=eQ'$ where
 $Q'=(g_X/g_Y)\tan
(\phi)$. Thus we see that the effect of the mixing of the
Stueckelberg term effectively changes $g_Y$ to $g_Y'=g_Y\cos\phi$.
Further, the photon also couples to the hidden sector fermions in
$J^\mu_X$ with a basic unit of charge $e'$, which in general will be 
irrational and small for a small mixing angle $\phi$. 
Thus one has the
interesting possibility of observing hidden sector particles which
carry small irrational charges through the Stueckelberg
phenomenon.

Next we discuss the couplings of the Z and Z$'$ bosons.\footnote{As an obvious remark, 
our construction preserves the GIM mechanism \cite{gim}.}
The couplings of the Z and Z$'$ bosons with the visible sector fields
read
\beqn
&& \hspace{-.8cm}
{\cal L}_{\rm NC} = \frac{g_2}{\cos(\theta)} \Big[
%\\
%&& \hspace{-1cm}
{\rm Z}_\mu \left(
 \cos(\psi) \left( \sin^2(\theta) Q J^\mu_{\rm em} -  J_2^{3\mu} \right) - \tan(\phi)\sin(\psi)\sin(\theta)
 \left( Q J^\mu_{\rm em} - J_2^{3\mu} \right) \right)
\non
&& \hspace{.7cm}
+~ {\rm Z}_\mu' \left(
\sin(\psi) \left( \sin^2(\theta) Q J^\mu_{\rm em} - J_2^{3\mu} \right) + \tan(\phi)\cos(\psi)\sin(\theta)
 \left( Q J^\mu_{\rm em} - J_2^{3\mu} \right) \right) \Big]\ . 
\nonumber
\eeqn
To estimate the range of parameters allowed by current experimental limits,  
it is instructive to compute the correction $\Delta$ to the Z boson mass 
relative  to that of the standard model, i.e.\ $M_{\rm
Z}=\frac{v}{2}\sqrt{g_2^2+g_Y^2}+\Delta=M_0+\Delta$,
where $M_0$ is the formula for the Z mass in the standard model at the tree level. 

The new parameters in the model beyond those
of the standard model are just $(M_1, M_2)$, but 
alternately, we can replace them by $(\Delta, M_{{\rm Z}'})$. 
Once these are fixed the mixing angles $\theta,\ \phi$ and $\psi$ can be computed. 
We do not undertake a global fit of electro-weak data in this paper, 
although this should be a worthwhile effort to determine the 
best values for $(\Delta, M_{{\rm Z}'})$. Rather here we demonstrate 
that one can pick specific parameters for the Stueckelberg sector, 
such that the fits of the standard model with the precision electro-weak
data are left essentially undisturbed, basically by arranging that the 
couplings between the Z$'$ and visible fermions are small.   
Thus, as an example, suppose we choose $|\Delta| =2\, $MeV and
$M_{{\rm Z}'}=150\, $GeV, where the choice of $|\Delta|$ is consistent
with the current error on the experimental determination
of $M_{\rm Z}$ \cite{pdg}. The above implies $\phi \sim 0.75^0$ and
$\psi \sim 0.18^0$ (via $M_1\sim 149\, {\rm GeV},\ M_2\sim 1.9\, {\rm GeV}$), 
and we use $\theta\sim \theta_W$.\footnote{Later we shall 
often neglect terms that are suppressed by extra factors of $\sin(\psi)$ 
compared to factors only suppressed by $\sin(\phi)$.} 
The smallness of these  angles  leaves 
the precision fits of the standard model with experiments intact. 
Typically, the couplings of extra gauge bosons are reduced relative to the couplings of the Z 
by a factor $M_{\rm Z}/M_{{\rm Z}'}$. 
There is, for instance, a vast literature on extra $U(1)$ gauge bosons in the
context of grand unified models such as $SO(10)$ or $E_6$
\cite{rizzo}, string and D-brane models \cite{cl2} and a
variety of other schemes \cite{march-russell,anoka}.
For models where the Z$'$ couples effectively with
the same strength as the Z, except for a Clebsch-Gordon coefficient,
one needs a large Z$'$ mass to achieve 
consistency with the precision electro-weak data \cite{rizzo,anoka}.
In the present example the couplings of the Z$'$ 
to quarks and leptons are suppressed by 
$M_2/M_1\sim 0.01$, which makes them small already.\footnote{The 
input to achieve this property was essentially to have the extra 
gauge boson $C_\mu$ not couple to 
standard model fermions.}
It is interesting to look for properties of the current model that lead to finite and 
potentially observable effects, such as quantites that depend on ratios of couplings. The 
cleanest signatures for Z$'$ would show up in resonant production in $e^+e^-$ collison at the 
Z$'$ mass peak. As we shall point out, one can then use the branching ratios for the different 
decay channels into $f\bar f$ species, and the forward-backward asymmetry in 
$e^+e^- \rightarrow \mu^+\mu^-$ as distinguishing features. 

The partial decay widths of the Z$'$ vector boson into visible fermions 
can easily be computed from rewriting the 
interactions of Z$'$ with fermions as 
\beqn \label{lrcpl}
{\cal L}^{\{ {\rm Z}'\}}_{\rm NC} = 
%{\rm Z}_\mu \Big[ g_L^f \bar f \gamma_\mu (1-\gamma_5) f 
%                                     + g_R^f \bar f \gamma_\mu (1+\gamma_5) f \Big] +  
     {\rm Z}'_\mu \sum_f \Big[ g_L^{f} \bar f \gamma^\mu (1-\gamma_5) f 
                                     + g_R^{f} \bar f \gamma^\mu (1+\gamma_5) f \Big] 
\eeqn
and using 
\beqn 
\Gamma ( {\rm Z} \rightarrow f\bar f ) = \frac{M_{{\rm Z}'}}{6\pi} \left( 
 (  g_L^f )^2 + (  g_R^f )^2 \right) + {\cal O}\left( \frac{M_f}{M_{{\rm Z}'}} \right) \  . 
\eeqn
>From the general formula for ${\cal L}_{\rm NC}$ we then obtain 
\beqn 
\Gamma ({\rm Z}'\rightarrow  l_i \bar l_i)
&=&
\frac{M_{{\rm Z}'}}{96\pi} \Bigg[ \Big( -\frac{g_2^2-g_Y^2\cos^2(\phi)}{\sqrt{g_2^2+g_Y^2\cos^2(\phi)}} 
 \sin(\psi) +g_Y\sin(\phi) \cos(\psi) \Big)^2 
\non
&& \hspace{1cm}
+~ 4\Big( \frac{g_Y^2\cos^2(\phi)}{\sqrt{g_2^2+g_Y^2\cos^2(\phi)}} 
  \sin(\psi) +g_Y\sin(\phi) \cos(\psi) \Big)^2 \Bigg] \ ,
\non 
\Gamma ({\rm Z}'\rightarrow  \nu_i \bar \nu_i)
&=& \frac{M_{{\rm Z}'}}{96\pi}
\Bigg[ \sqrt{g_2^2+g_Y^2\cos^2(\phi)} \sin(\psi) +g_Y\sin(\phi) \cos(\psi) \Bigg]^2 \ , 
\non
\Gamma ({\rm Z}'\rightarrow  d_i \bar d_i)
&=&
\frac{M_{{\rm Z}'}}{4\pi}
(g_2^2+g_Y^2\cos^2(\phi))
\Bigg[ \Big( \Big( -\frac{1}{4} +\frac{1}{3} \sin^2(\theta) \Big) \sin(\psi) 
\non
&&
\hspace{-1.2cm}
+~ \frac{1}{12}  \sin(\theta)\tan(\phi) \cos(\psi) \Big)^2
+ \Big( \frac{1}{4} \sin(\psi) +\frac{1}{4}\sin(\theta) \tan(\phi) \cos(\psi) \Big)^2 \Bigg] \ , 
\non
\Gamma ({\rm Z}'\rightarrow  u_i \bar u_i)
&=&
\frac{M_{{\rm Z}'}}{4\pi}
(g_2^2+g_Y^2\cos^2(\phi))
\Bigg[ \Big( \Big( \frac{1}{4} -\frac{2}{3} \sin^2(\theta) \Big) \sin(\psi) 
\\ 
&&
\hspace{-1.7cm} 
-~ \frac{5}{12}  \sin(\theta)\tan(\phi) \cos(\psi) \Big)^2
+\Big( 
-\frac{1}{4} \sin(\psi) -\frac{1}{4}\sin(\theta) \tan(\phi) \cos(\psi)\Big)^2
\Bigg] \ . 
\nonumber 
\eeqn
Further, the couplings of Z$'$ to W$^+$W$^-$, ZZ, and Higgs fields are
proportional to $\sin(\psi)$, since $A_\mu^3 = - \cos(\theta)\sin(\psi) {\rm Z}'+\,\cdots$.   
In the limit of small $\psi$ we neglect these contributions, and replace $\cos(\psi)\sim 1$. 
With this simplification, the total decay width below or above 
the $t\bar t$ threshold is given by
\beqn \label{width}
\Gamma ({\rm Z}'\rightarrow \sum_i f_i \bar f_i)
= M_{{\rm Z}'}\, g_Y^{2} \sin^2(\phi) \times 
{{\frac{103}{288\pi}~~ {\rm for}~~ M_{{\rm Z}'}<2m_t}\brace{\frac{5}{12\pi}~~ {\rm for}~~ M_{{\rm Z}'}>2m_t}} \ . 
\label{totalwidth}
\eeqn
For $M_{{\rm Z}'}=150$ GeV, the total decay width lies in the range
$0.6-80\,$MeV for $\phi\sim 1^0-10^0$, while for
$M_{{\rm Z}'}=1\,$TeV it would lie in the range $4\, {\rm MeV} - 0.6\, {\rm GeV}$. 
This is to be contrasted with Z$'$
widths of up to $100\, $GeV or above for a Z$'$ boson with mass of $1-2\, $TeV, as in
most other models \cite{anoka}. Thus the observation of a  sharp Z$'$ could be seen as 
potential signal for a Stueckelberg gauge boson. 

An interesting phenomenon relates
to the possibility of observing the hidden sector via the
Z$'$ decay. In the analysis of Eq.~(\ref{totalwidth})
we have ignored the couplings of the Z$'$ to the hidden sector
fields in the current $J_X^\mu$. If the masses of the hidden sector fields are smaller 
than $M_{{\rm Z}'}$, there would be
additional contributions to the decay width of Z$'$.
Thus if kinematically allowed, the Z$'$ would decay into a pair
of hidden sector fermions with small irrational electric charges $Q'e$.
Because of the small charges of such particles, we expect that they would typically
escape detection with the current sensitivity of detectors and would appear
as electrically neutral. It should be challenging to find ways to detect
such particles. 
We note that hidden matter which is neutral under $SU(2)_L\times U(1)_Y$ would still
possess charges in units of $e'$, even  if $C_{\mu}$ also
coupled to the visible sector fields. However, in this case, 
the result of Eq.~(\ref{width}) would be modified, and 
the Z$'$ width will be broadened \cite{kn}.

>From the above, one can easily get the branching ratios of Z$'$ into different 
fermionic modes, which will have distinctive characteristics. For the decay of Z$'$
into up quarks and down quarks one finds 
$R_{u/d}^{{\rm Z}'}={{\rm BR}({\rm Z}'\rightarrow u_i\bar u_i)}/
{{\rm BR}({\rm Z}'\rightarrow d_i\bar d_i)} \sim {17}/{5}$.
This is to be contrasted with the result for the Z decay
in the standard model where $R_{u/d}^{\rm Z}\sim 0.78$. Similarly,
one has 
$R_{l/\nu}^{{\rm Z}'}= {{\rm BR}({\rm Z}'\rightarrow l_i\bar l_i)}/
{{\rm BR}({\rm Z}'\rightarrow \nu_i\bar \nu_i)} \sim 5$, 
compared to $R_{l/\nu}^{\rm Z}\sim 0.5$, and for the quarks consider e.\ g.\ 
$R_{b/\tau}^{{\rm Z}'}$=${{\rm BR}({\rm Z}'\rightarrow b\bar b)}/$
${{\rm BR}({\rm Z}'\rightarrow \tau^+\tau^-)} \sim {1}/{3}$, while 
$R_{b/\tau}^{\rm Z}\sim 4.4$.
Another  ratio of interest is 
$R_l^{{\rm Z}'}$=${\rm BR} ({\rm Z}'\rightarrow {\rm had})/$
${\rm BR} ({\rm Z}'\rightarrow l^+l^-)$ $\sim 49/15$, below the $t\bar t$ threshold, 
and $22/5$, above, compared to $R_l^{\rm Z}\sim 20.77$, and finally 
${\rm BR}({\rm Z}'\rightarrow \sum_il_i^+l_i^-)\sim 44\%$ for below, and $38\%$ 
above the $t\bar t$ threshold. For the Z boson one has in the standard model 
${\rm BR}({\rm Z}\rightarrow \sum_il_i^+l_i^-)\sim 10\%$. 
These branching ratios of the ${\rm Z}'$ 
are in fact drastically different from the decay branching
ratios of the Z$'$ in many other models \cite{rizzo,anoka}.

Another important signature for the Stueckelberg Z$'$ is the
forward-backward asymmetry $A_{\rm FB}$ in $e^+e^-\rightarrow \mu^+\mu^-$
experiment. The forward-backward asymmetry parameter $A_{\rm FB}$
is defined by 
$A_{\rm FB}=({\int_{0}^1 dz \frac{d\sigma}{dz} -\int_{-1}^0
dz \frac{d\sigma}{dz}})/{(\int_{-1}^1 dz \frac{d\sigma}{dz})}$.
At the Z$'$ pole it can be approximated by 
$A_{\rm FB}(s=M^2_{{\rm Z}'}) \sim (3/4){((g_L^{l})^2-(g_R^{l})^2)^2}/
{((g_L^{l})^2+(g_R^{l})^2)^2}$, with universal couplings 
$g_{L,R}^e=g_{L,R}^\mu=g_{L,R}^l$, 
as defined through (\ref{lrcpl}), and where we have only taken contributions due 
to Z$'$ exchange into account. One finds
\beqn \hspace{.5cm}
\frac{g_R^l}{g_L^l} = 2 \frac{(1+\delta)\tan^2(\theta)}{(1+\delta)\tan^2(\theta)-1} \ , 
\quad 
\delta = \frac{\tan(\phi)}{\sin(\theta)\tan(\psi)} = \frac{M_{{\rm Z}'}^2-M_{\rm Z}^2}{M_{\rm Z}^2-M_{\rm W}^2} + {\cal O}(M_2^2) \ . 
\nonumber 
\eeqn
We then have $2\le {g_R^l}/{g_L^l}< 3.3$ for $M_{{\rm Z}'}>140\,$GeV, 
and ${g_R^l}/{g_L^l}$ goes to $2$ asymptotically, which is its value for small $\psi$.
For ${g_R^l}/{g_L^l} \sim 2$ one has $A_{\rm FB}(s=M^2_{{\rm Z}'}) \sim 0.27$.
The asymmetry at the Z pole on the contrary is $A_{\rm FB}(s=M^2_{{\rm Z}})\sim 0.02$. 
Also of interest is the quantity 
$R_{\rm peak}^{{\rm Z}'}(\mu^+\mu^-)=\sigma (e^+e^-\rightarrow \mu^+\mu^-)/
\sigma (e^+e^-\rightarrow \mu^+\mu^-)_{\rm QED}$ at the Z$'$ pole
and similarly the quantity $R_{\rm peak}^{{\rm Z}'}({\rm charged~lep+ had})=
\sigma(e^+e^-\rightarrow {\rm charged~lep+had})/
\sigma (e^+e^-\rightarrow \mu^+\mu^-)_{\rm QED}$ at the Z$'$ pole.
Our analysis gives $R_{\rm peak}^{{\rm Z}'}(\mu^+\mu^-)=9/(4\alpha^2)\sim 3.5\times 10^{4}$,  
and $R_{\rm peak}^{{\rm Z}'}({\rm charged~lep+had})=333/(20\alpha^2)\sim 2.7 \times 10^{5}$. 
This is to be compared with 
$R_{\rm peak}^{\rm Z}({\rm charged~lep+had})\sim 0.4\times10^{4}$. 
It would be interesting to see if experiment can reach sensitivity necessary
to test the presence of sharp resonances in the region of the Z$'$ mass.

The phenomenology of the extended model at the hadron colliders and 
in cosmology should also be of interest  and needs investigation. 
There are of course many avenues for further generalizations and
 modifications of our model, 
including for instance a Stueckelberg extension of the minimal supersymmetric standard model.
This will be discussed elsewhere \cite{kn}.
\\
%[-.3cm]

\noindent {\bf Acknowledgements}\\ 
Useful conversations with Stephen Reucroft and Darien Wood regarding
experiments are acknowledged.
The work of B.~K.~was supported by the German Science Foundation (DFG) and in part by
funds provided by the U.S. Department of Energy (D.O.E.) under cooperative research agreement
$\#$DF-FC02-94ER40818.  The work of P.~N. was supported in part by
the U.S. National Science Foundation under the grant NSF-PHY-0139967\\

%\clearpage

\end{document}